\documentclass[journal]{IEEEtran}
\usepackage{graphicx}
\usepackage{graphics} 
\usepackage{epsfig} 
\usepackage{mathptmx} 
\usepackage{times} 
\usepackage{amsmath} 
\usepackage{amssymb}  
\usepackage[T1]{fontenc}
\usepackage{bm}
\usepackage{xcolor}
\usepackage[fitting]{tcolorbox}
\usepackage{cite}
\usepackage{hyperref}
\usepackage[utf8]{inputenc}
\usepackage[colorinlistoftodos]{todonotes}
\usepackage{multirow}
\usepackage{soul}
\setstcolor{red}

\usepackage{caption}
\begin{document}

\title{\LARGE \bf A statistical reconstruction algorithm for positronium lifetime imaging using time-of-flight positron emission tomography
}
\author{Hsin-Hsiung Huang, Zheyuan Zhu,  Slun Booppasiri, Zhuo Chen, Shuo Pang, and Chien-Min Kao
\thanks{This work did not involve human subjects or animals in its research.}
\thanks{This work was partially supported by NSF grant DMS-1924792 (Huang), DMS-2318925 (Huang)
and NIH grant R01-EB029948 (Kao). Corresponding author: Hsin-Hsiung Huang.}
\thanks{Z. Zhu and S. Pang are with CREOL, The College of Optics and Photonics, University of Central Florida, Orlando, FL 32816 (e-mails: zyzhu@knights.ucf.edu,  pang@ucf.edu)}
\thanks{C.-M. Kao is with Department of Radiology, University of Chicago, Chicago, IL 60637 (e-mail: ckao95@uchicago.edu).}
\thanks{Z. Chen is with Department of Mathematics, University of Arizona, Tucson, AZ 85721 (e-mail: zchen1@math.arizona.edu).}
\thanks{S. Booppasiri (email: slun.booppasiri@ucf.edu) 
 and H.-H. Huang are with Department of Statistics and Data Science, University of Central Florida, Orlando, FL 32816 (email: hsin.huang@ucf.edu)}
}


\maketitle

\begin{abstract}
Positron emission tomography (PET) is an important modality for diagnosing diseases such as cancer and Alzheimer's disease, capable of revealing the uptake of radiolabeled molecules that target specific pathological markers of the diseases. Recently, positronium lifetime imaging (PLI) that adds to traditional PET the ability to explore properties of the tissue microenvironment beyond tracer uptake has been demonstrated with time-of-flight (TOF) PET
and the use of non-pure positron emitters.
However, achieving accurate reconstruction of lifetime images from data acquired
by systems having a finite TOF resolution still presents a challenge.
This paper focuses on the two-dimensional PLI, introducing a maximum likelihood estimation (MLE) method that employs an exponentially modified Gaussian (EMG) probability distribution
that describes the positronium lifetime data produced by TOF PET. We evaluate the performance of our EMG-based MLE method against approaches using exponential likelihood functions and penalized surrogate methods. Results from computer-simulated data reveal that the proposed EMG-MLE method can yield quantitatively accurate lifetime images.
We also demonstrate that the proposed MLE formulation can be extended to handle
PLI data containing multiple positron populations.
\end{abstract}
\IEEEoverridecommandlockouts
\IEEEpeerreviewmaketitle

\section{Introduction}
\label{sec:intro}
Positronium lifetime imaging (PLI) with time-of-flight (TOF) positron emission tomography (PET) represents a recent advancement in medical imaging, with its feasibility substantiated through various experimental studies \cite{moskal2019feasibility, moskal2019positronium, moskal2020performance, shibuya2020oxygen, zgardzinska2020studies}. Traditional PET is known for assessing the functional state of organs or tissues through the uptake of specific PET molecules. In contrast, PLI aims to measure the lifetime of positronium that is a meta-stable electron-positron pair formed by the positrons emitted by PET molecules \cite{harpen2004positronium}. More precisely, it measures the lifetime of so-called ortho-positronium (o-P) that can be
significantly affected by the interaction of an o-P with nearby molecules, such as oxygen,
that possess an unpaired electron. As a result, o-P lifetime may serve as a quantitative marker of such molecule presences in the tissue microenvironment, independent of the traditional PET molecule uptake mechanisms. This capability can be significant in clinical contexts. For instance, it may be used to identify hypoxic tissues that are often resistant to therapies \cite{shibuya2020oxygen,moskal2021positronium2}, potentially enabling enhanced treatment strategies for conditions like cancer. 

Today's TOF-PET systems have a coincidence resolving time (CRT) ranging from 200 to 600 ps full width at half maximum (FWHM) \cite{karp2020pennpet, spencer2021performance, alberts2021clinical}, corresponding to a spatial uncertainty of 3-9 cm. Consequently, the measurement can include a mixture of events originating from different locations with varying lifetimes. Initial experimental implementations of PLI address this issue by placing samples sufficiently apart relative to the system's TOF resolution \cite{moskal2021positronium, moskal2021testing}. However, in practical applications, such spatial separation is often not feasible.

The inverse Laplace transform method \cite{shibuya2022using} has been proposed to separate various components in a lifetime histogram. This method significantly reduces information loss by accurately decomposing overlapping lifetime signals, which is a common issue with traditional averaging techniques. Unlike conventional methods that average lifetime data and obscure fine details, the inverse Laplace transform method preserves the distinct lifetime components, leading to a more precise and detailed analysis of the data. However, it does not perform image reconstruction.
Several image reconstruction methods for PLI with TOF-PET have been recently reported \cite{qi2022positronium,huang2024statistical,huang2023high,shopa2023positronium,chen2023properties}. Qi and Huang \cite{qi2022positronium} developed a penalized surrogate (PS) method to produce a regularized maximum likelihood (ML) solution. Their model is based on a single-exponential probabilistic model for the PLI data, which does not account for the effects of finite CRT on the lifetime measurement. This paper introduces an extended statistical model for 2-dimensional TOF-PET PLI data that allows the lifetime distribution to contain multiple components and considers the exponentially-modified Gaussian distribution for each component to account for the uncertainty in time measurement.
The ML solution according to the model was solved using the Limited-memory Broyden-Fletcher-Goldfarb-Shanno Bound (L-BFGS-B) method from scipy.optimize \cite{SciPy-NMeth2020}, imposing nonnegativity condition for the solution. 

The remainder of this paper is organized as follows: Section II details the formulation of the statistical model for the PLI list-mode data and the derivation of an algorithm to obtain the maximum likelihood estimates of lifetime images. Section III describes the computer-simulation study and presents the results. Section IV offers a summary and conclusions.

\section{Statistical model for TOF-PET PLI data}

\subsection{Detection of a PLI event with TOF-PET}
\label{sec::detection_of_PLI_event}

\begin{figure}[t]
\centering
\includegraphics[width=8cm]{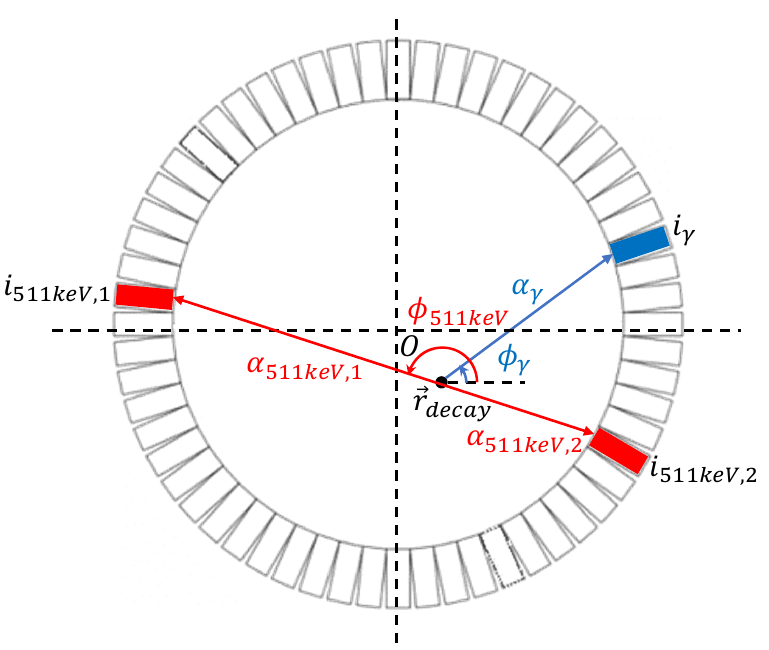}
\caption{Detection of a PLI event using a 2D TOF-PET system.}
\label{fig:simu_setup}
\end{figure}

Fig.~\ref{fig:simu_setup} illustrates a two-dimensional (2D) TOF-PET system comprising a ring of uniformly spaced detectors that are numerically labeled by an integer $i$.
It is assumed that a non-pure positron-emitter such as Sc-44, which emits a positron and a gamma ray effectively at the same time (called a prompt gamma), is employed.
Compared to the popular PET isotopes like F-18, many non-pure positron-emitters
have a larger positron range\footnote{Na-22 is a notable exception. See \cite{Jodal_2014}.}
that can limit the resulting image resolution\cite{matulewicz2021radioactive,choinski2021prospects, ferguson2019comparison}. 
In this work, we shall neglect the positron range for it does not affect the fundamental validity of the proposed reconstruction method \cite{qi2022positronium}.
Similarly, photon acolinearity is also neglected \cite{shibuya2007annihilation}.

The decay of an isotope like Sc-44 releases a prompt gamma and a positron
at location ${\bm r}_{decay}$ and time $t_{decay}$.
As illustrated in  Fig.~\ref{fig:simu_setup},
the prompt gamma travels a distance $\alpha_\gamma$ towards the detector ring at angle $\phi_\gamma$, captured by detector $i_\gamma$ at time $t_\gamma = t_{decay} + \alpha_\gamma/v_c$ where $v_c$ is the speed of light.
The positron can momentarily form a positronium before it annihilates with an electron.
The elapse time before annihilation $\tau$ is described by an exponential distribution 
\cite{moskal2021positronium}, given by
\begin{equation}
p(\tau;\lambda)=\mbox{Exp}(\tau;\lambda)=\begin{cases}
    \lambda e^{-\lambda \tau} & \tau\geq 0\\
    0                         & \tau<0.
\end{cases},
\label{lifetime_exp_dist}
\end{equation}
Lifetime refers to the inverse of the rate constant $\lambda$;
these two terms will be used interchangeably in this paper.
There are, in fact, two positronium populations:
para-positronium (p-P) and ortho-positronium (o-P).
Positrons that do not form positroniums before annihilation, called
direct annihilation (DA), also can exist for a finite time.
The DA and p-P lifetimes are substantially shorter than the o-P lifetime.
Unlike the o-P lifetime, the DA and p-P lifetimes,
and the proportions of the DA, p-P and o-P populations,
are not sensitive to the environment.
Therefore, the main interest in PLI is the o-P lifetimes.
Please see \cite{shibuya2020oxygen,moskal2021positronium} for in-depth discussion.

On annihilation, as depicted by the red line in the Fig.~\ref{fig:simu_setup}, two opposite 511\,keV gamma rays are created.
They travel from ${\bm r}_{decay}$ at a random angle $\phi_{511keV}$ with respect to $\phi_\gamma$
and are detected by
detectors $i_{511keV,1}$ and $i_{511keV,2}$
at time 
$t_{511keV,1}=t_{decay}+\tau+\alpha_{511keV,1}/v_c$
and $t_{511keV,2}=t_{decay}+\tau+\alpha_{511keV,2}/v_c$
respectively,
where $\alpha_{511keV,1}$ and $\alpha_{511keV,2}$
are the distances that they travel.

The conventional TOF-PET system reports $i_{511keV,1}$, $i_{511keV,2}$,
and the TOF given by
\begin{equation}
\Delta t_{511keV} = 
t_{511keV,1}-t_{511keV,2}=(\alpha_{511keV,1}-\alpha_{511keV,2})/v_c.
\label{eq::TOF}
\end{equation}
We assume that the system is extended to be capable of triple-coincidence detection and reports additionally $i_{\gamma}$ and
\begin{align}
\Delta t_\gamma &= (t_{511keV,1}+t_{511keV,2})/2-t_\gamma \nonumber\\
&= \tau+(\alpha_{511keV,1}+\alpha_{511keV,2}-2\alpha_\gamma)/(2v_c).
\label{eq::delta_T_gamma}
\end{align}
Note that $\alpha_{511keV,1}+\alpha_{511keV,2}$ 
can be determined from the locations of detectors
$i_{511keV,1}$ and $i_{511keV,2}$.
Additionally, if $\Delta t_{511keV}$ is precisely known,
${\bm r}_{decay}$ can be identified,
and then $\alpha_\gamma$ can be computed
from $i_\gamma$ and ${\bm r}_{decay}$.
Then, Eq.~(\ref{eq::delta_T_gamma}) can be used
to compute $\tau$ from $\Delta t_\gamma.$

In a real system, time measurement has limited precision
and is typically binned and stored as integers.
CRT refers to 
the uncertainty of $\Delta t_{511keV}$ in FWHM.
With a finite CRT,
${\bm r}_{decay}$ cannot be precisely determined.
A CRT of 200 ps to 600 ps translates to 
an uncertainty of 3 cm to 9 cm in ${\bm r}_{decay}$.
Similarly, $\Delta t_\gamma$ will have limited precision
and be binned.
Hence, in Eq.~(\ref{eq::delta_T_gamma}) $\alpha_\gamma$
is not precisely observed and 
all the time measurements involved contain
statistical variations.

\subsection{Probability model for the TOF-PET PLI list-mode data}

In conventional PET, data is associated with a line of response (LOR)
$\mathcal{L}(i_{511keV,1},i_{511keV,2})$
connecting two detectors $i_{511keV,1}$ and $i_{511keV,2}$ 
that capture the annihilation photons. 
In TOF-PET, an LOR is further segmented into a number of non-overlapping segments referred to as lines of segment (LOS), using a TOF binsize that equals to $\text{CRT}/2$.
Each LOS is identified by a multi-index
$c = (i_{511keV,1},i_{511keV,2},m)$, indicating the specific TOF bin $m$ on the LOR. 
We assume that the PLI events are given by
$w=(c,i_\gamma$, $\Delta t_\gamma)$, where \( c \) identifies the LOS, 
\( i_\gamma \) is the detector that captures the prompt gamma,
and \( \Delta t_\gamma \) is the time difference between the detection of the
annihilation photons and the prompt gamma.
These PLI events are stored as list-mode (LM) data, represented as \( \mathcal{W}_{N_k} = \{w_k\}_{k=1}^{N_k} \), where \( k \) indexes a list of
event words $w_k=(c_k,i_{\gamma,k},\Delta t_{\gamma,k})$
and \( N_k \) is the total number of events.

\subsubsection{Calculation of the system matrix}

Each element \( H_{c,j} \) of the system matrix \( \bm{H} \) is designed to reflect the likelihood that an annihilation taking place in image pixel \( j \) results in a detection at LOS \( c \). The system matrix is calculated using a ray-tracing method \cite{siddon1985fast,zhu2019photon}, identifying which pixels are intersected by \( \mathcal{L}(i_{511keV,1},i_{511keV,2}) \) and the boundary points of intersection. A Gaussian function, whose width is given by the CRT, is placed along \( \mathcal{L}(i_{511keV,1},i_{511keV,2}) \) and centered at the midpoint of these boundaries for each intersecting pixel \( j' \). The integral of this Gaussian over each TOF bin yields the values for \( H_{(i_{511keV,1},i_{511keV,2},m),j'} \), using 
$(2/\sqrt{\pi})\int_a^b e^{-t^2} dt = \textrm{erf}(b)-\textrm{erf}(a)$, where $\textrm{erf}(x)$ is the error function. Pixels not intersected by the LOS are assigned a zero value in the system matrix.

\subsubsection{Maximum likelihood estimation}
Let the vectors $\bm{f}=[f_j]$ and $\bm{\lambda}=[\lambda_j]$
be, respectively, the activity and o-P rate-constant images,
where $f_j\geq 0$ and $\lambda_j\geq 0$ are their respective
values in voxel $j$.
$\tau_k$ can be computed from $c_k$, $i_{\gamma,k}$, and $\Delta t_{\gamma,k}$
to produce the LM data set
$\mathcal{W}^0_{N_k}=\{w_k^0\}_{k=1}^{N_k}$ with $w_k^0 = (c_k, \tau_k)$
in place of $\mathcal{W}_{N_k}$.
We will derive in the Appendix that 
the log-likelihood for $\bm{\lambda}$ given $\bm{f}$ and $\mathcal{W}^0_{N_k}$ is
\begin{equation}
\ell(\bm{\lambda}; \bm{f},\mathcal{W}^0_{N_k}) =\sum_{k=1}^{N_k}\log
\left(\sum_{j=1}^{N_j}
H_{c_k,j}f_j\,\mbox{EMG}(\tau_k;\lambda_j,\sigma^2)\right),
\label{PET::profile_log_likelihood_PLI}
\end{equation}
where $\text{EMG}(\tau; \lambda, \sigma^2)$, the exponentially modified Gaussian (EMG) distribution, is the convolution of $\mbox{Exp}(\tau;\lambda)$ with
a zero-mean Gaussian with standard deviation $\sigma$ that is
introduced to take in account the uncertainty in measurement of $\tau$.
The convolution can be explicitly evaluated to yield \cite{grushka1972characterization}:
\begin{equation}\small
    \mbox{EMG}(\tau;\lambda,\sigma^2)=\frac12\lambda e^{-\lambda(\tau-\frac12\sigma^2\lambda)}\left(1+\mbox{erf}\left(\frac{\tau-\lambda\sigma^2}{\sqrt{2}\sigma}\right)\right),
\label{eq:EMG}
\end{equation}
where $\mathrm{erf}(x)$ is the error function.
It can be checked that $\mbox{EMG}(\tau;\lambda,0)=\mbox{Exp}(\tau;\lambda)$

Note that the TOF-PET PLI LM data contains the
traditional TOF-PET LM data $\mathcal{C}_{N_k} = \{c_k\}_{k=1}^{N_k}$.
Approaches for solving $\bm{f}$ from $\mathcal{C}_{N_k}$, and
from histogram-mode data derived from it, are well established \cite{leahy2000statistical}.
A popular algorithm is the ordered-subsets (OS) expectation-maximization (EM)
algorithm \cite{hudson1994accelerated} that is an accelerated version of
the EM algorithm for PET image reconstruction that can produce
ML estimation (MLE) for $\bm{f}$.
In this paper, we will employ the OS-EM algorithm for estimation of $\bm{f}$.
The MLE of $\bm{\lambda}$ is then obtained by maximizing 
$\ell(\bm{\lambda};\bm{f},\mathcal{W}_{N_k}^{0})$ given in Eq.~(\ref{PET::profile_log_likelihood_PLI}) by using either the true $\bm{f}$
or its estimate $\bm{f}$ obtained by OS-EM.
Maximization is carried out by using the Limited-memory Broyden-Fletcher-Goldfarb-Shanno Bound (L-BFGS-B) method available from scipy.optimize \cite{SciPy-NMeth2020}.
Positivity condition on $\bm\lambda$ is implemented but no explicit noise regularization schemes are introduced.

\section{Computer-simulation studies}
\label{sec:simu}

\subsection{Data generation}

We generated the TOF-PET PLI LM data, $\mathcal{W}_{N_k}$, using Monte Carlo methods tailored for a scanner configuration with $N_{det}$ detectors uniformly distributed on a diameter $D$. Given $\bm{f}$ and $\bm{\lambda}$, we simulated the decay process as follows.

First, given a desired number of total decays, the image $\bm{f}$ was scaled such that its
pixel values represented the desired mean number of decays.
Then, a list of decays was sampled from a Poisson distribution defined by these means.
For each decay, its position,
${\bm r}_{decay} = (x_{decay}, y_{decay})^T$,
was sampled within the bounds of the pixel area
$A_j = [x_j - \Delta x/2, x_j + \Delta x/2) \times [y_j - \Delta y/2, y_j + \Delta y/2)$,
where $(x_j, y_j)$ represented the center coordinates of pixel $j$, and
$\Delta x$ and $\Delta y$ were the pixel dimensions along the $x$ and $y$ axes, respectively, according to a uniform distribution, $\mathcal{U}_{A_j}$.

Then, a prompt gamma emitted in a random angle $\phi_{\gamma}$, sampled from $\mathcal{U}{[0,2\pi)}$, was generated at ${\bm r}_{decay}$.
The travel distance $\alpha_{\gamma}$ before detection was determined by solving:
\begin{equation}
\label{eq:alpha_decay}
|{\bm r}_{decay}+\alpha_\gamma\hat{\bm \phi}_\gamma|= D/2,
\end{equation}
where $\hat{\bm \phi} = (\cos \phi, \sin \phi)^T$ is the unit direction vector. 
This equation has two solutions given by
\begin{equation}
\alpha_\gamma^{\pm} = -\hat{\bm \phi}_{\gamma}^T {\bm r}_{decay}
\pm \sqrt{(\hat{\bm \phi}_{\gamma}^T{\bm r}_{decay})^2
- ||{\bm r}_{decay}||^2 + D^2/4},
\label{eq::travel_distance}
\end{equation}
due to uniform sampling of $\phi_{\gamma}$
the solution $\alpha_\gamma^+$ 
can be arbitrarily used for consistency,
yielding $\bm{r}_{detect,\gamma}=\bm{r}_{decay}+\alpha_\gamma^+ \hat{\bm \phi}_{\gamma}$.
The index $i_\gamma$ of the detector that
the prompt gamma hits was calculated by:
\begin{equation}
i_\gamma =\left\lfloor \left(\frac{N_{det}}{2\pi}\right)
\measuredangle{\bm r}_{detect,\gamma} \right\rfloor,
\label{eq::hit_detector}
\end{equation}
where $\measuredangle {\bm r}$ denotes the angle of vector $\bm r$ in polar coordinates, and $\lfloor x \rfloor$ represents the floor function.

Similarly, two opposing annihilation photons were emitted
at $\bm{r}_{decay}$ in a random angle $\phi_{511keV}$ sampled from $\mathcal{U}{[0,2\pi)}$. The events were assumed to occur after an elapsed time drawn from an appropriate distribution. When only the o-Ps population was considered, this distribution was a single exponential distribution with a rate constant $\lambda_j$ defined for pixel $j$.
When two populations were considered, the distribution was a weighted mixture of two exponential distributions, each with its own rate constant defined for pixel $j$ (see Sect.~\ref{subsect:two-population}).
The distances $\alpha_{511keV,1}$ and $\alpha_{511keV,2}$,
and detector indices $i_{511keV,1}$ and $i_{511keV,2}$
were calculated as described above,
now using both solutions in Eq.~(\ref{eq::travel_distance}).
The emission angles $\phi_\gamma$ and $\phi_{511keV}$ were independent.
The detection time $t_{511keV,1}$, $t_{511keV,2}$, and $t_{\gamma}$ relative to the decay time $t_{decay}$ were readily calculated from their travel distances divided by $v_c$, and the elapse time between positron annihilation and isotope decay.

To account for uncertainty in time measurement, every detection time was perturbed by a random number drawn from a Gaussian distribution having zero mean and standard deviation (SD) $\sigma_1$. 
By convention, the CRT of the a TOF-PET system is the FWHM uncertainty in the coincidence time measurement. Therefore, $\sigma_1 = (\text{CRT}/2\sqrt{2\ln 2})/\sqrt{2}$.
The TOF was then calculated using $\Delta t_{511keV}=t_{511keV,1}-t_{511keV,2}$, and
\begin{equation}
\Delta t_\gamma = (t_{511keV,1}+t_{511keV,2})/2-t_\gamma.
\label{eq:delay-time}
\end{equation}
The measured $\tau$ should be computed from $\Delta t_\gamma$
by correcting for the travel-time difference
between the prompt gamma and annihilation photons.
This travel time difference can be estimated from 
$i_{511keV,1}$, $i_{511keV,2}$, $i_\gamma$, and $\Delta t_{511keV}$.
For simplicity, in this work we used the exactly known travel distances
and computed $\tau$ as:
\begin{equation}
    \tau = \Delta t_{\gamma}-\frac{
    \alpha_{511keV,1}+\alpha_{511keV,2}-2\alpha_{\gamma}}{2v_c}.
\label{measured_tau}
\end{equation}

\begin{figure}[t]
\centering
\includegraphics[scale=0.58]{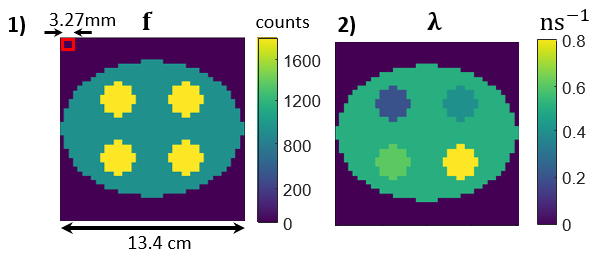}
\caption{(1) The activity image and
(2) the rate-constant image of Phantom 1.
These images consists of
3.27$\times$3.27~mm\textsuperscript{2}
square pixels. Radius of each disc is 12 mm.}
\label{phantoms}
\end{figure}

\begin{figure}
    \centering
    \includegraphics[width=0.5\textwidth]{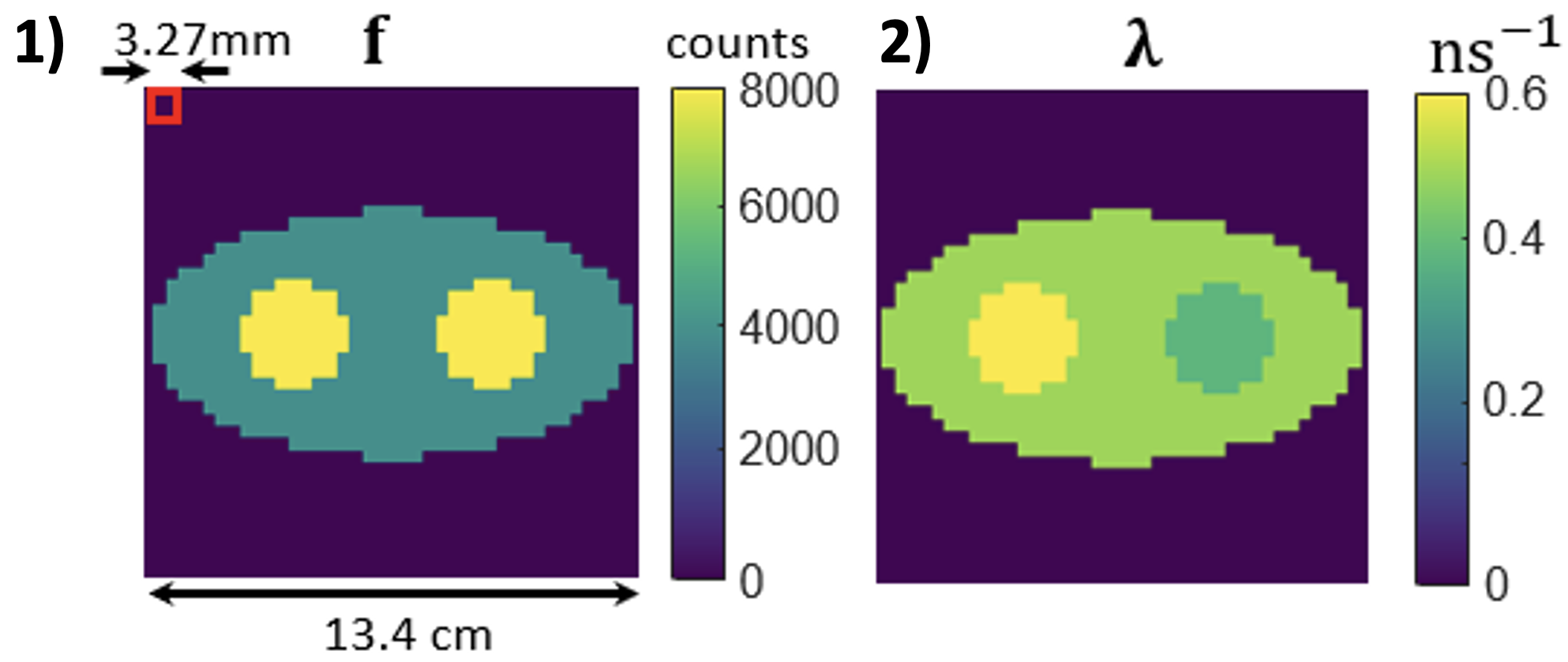}
    \caption{(1) The activity image and (2) the rate-constant image of Phantom 2, with a pixel size of 3.27 mm.
    This phantom was used to examine the reconstruction method under different CRTs.}
      \label{fig:nmse_mixed}
\end{figure}

We generated TOF-PET PLI LM data with $N_{det} = 364$, $D = 57.2$ cm, and a specified CRT.
Two numerical phantoms were considered, each consisting of an activity image and a rate-constant image.
The activity and rate-constant images of Phantom 1 in Fig.~\ref{phantoms} both contain four discs (radii = 12mm) on top of a circular background.
In terms of pixel index, the centers of the upper left, upper right, lower left and lower right discs are at (13.5,14.5), (28.5,14.5), (13.5,27.5), and (28.5, 27.5) respectively.
For the activity image, the disc-to-background ratio is 2:1.
For the rate-constant image, the four discs have different $\lambda$ values
(0.2, 0.4, 0.6, and 0.8 ns$^{-1}$) and the background value is 0.5 ns$^{-1}$.
Phantom 2 shown in Fig.~\ref{fig:nmse_mixed} was considered for examining the effects of CRT. Its activity and rate-constant images feature an elliptical background enclosing two discs. In Phantom 2, the activity ratio is also 2:1. The decay-rate image for o-Ps includes two discs with different decay constants, $\lambda$ (0.4 and $0.6$ ns$^{-1}$), in contrast to the background ellipse with $\lambda = 0.5$ ns$^{-1}$.
Both phantoms contained 41$\times$41 square pixels measuring 3.27$\times$3.27 mm\textsuperscript{2}.

Unless mentioned otherwise, the expected total number of events to generate in each simulation run was one million. Our simulations considered only valid triple-coincidence events; attenuation, scattering, and random events were not included.

\subsection{PLI reconstruction}

\subsubsection{EMG/EXP-MLE methods}
Below, the proposed PLI reconstruction method is referred to as EMG-MLE.
When estimating $\tau$ using Eq.~\eqref{measured_tau}, the variance in $\tau$ equals to the variance in $\Delta{t}_\gamma$, which, according to Eq.~\eqref{eq:delay-time}, is given by $(3/2)\sigma_1^2$. Therefore, the $\sigma$ parameter in the EMG model was set to $\sqrt{3/2}\sigma_1=(\sqrt{3}/2)(\text{CRT}/2.35)$. We may also set $\sigma=0$. In this case, the EMG distribution reduces to the Exp distribution. For distinction, reconstruction using $\sigma=0$ is referred to as Exp-MLE.

\subsubsection{The reference method}

The penalized surrogate (PS) method developed by Qi and Huang \cite{qi2022positronium} was considered for comparison. This method
 updates an estimate for $\bm{\lambda}$ by:
\begin{equation}
\label{penalized_surrogate}
\hat{\lambda}_j^{(m+1)} = \frac{\sum_{k=1}^{N_k} u_{c_k,j}^{(m)}}{\beta_j^m + \sum_{k=1}^{N_k} u_{c_k,j}^{(m)}\tau_k},
\end{equation}
where $m$ is the iteration number,
\begin{equation}
u_{c_k,j}^{(m+1)} = 
\frac{H_{c_k,j} \hat{f}_j \lambda_j^{(m)} \mbox{exp}(-\tau_k\times \lambda_j^{(m)})}
{\sum_{l=1}^{N_j} H_{c_k,l} \hat{f}_l \lambda_l^{(m)} \mbox{exp}(-\tau_k\times\lambda_l^{(m)})},
\label{eq:PS-2}
\end{equation}
and $\beta_j^m = \eta_j^m\beta$ where $\beta\geq 0$ is a regularization
parameter and $\eta_j^m$ is calculated from neighbors of pixel $j$ to
adjust the strength of regularization locally.

\subsubsection{Initial estimate and selection criterion}
As already mentioned above, either the known activity image $\boldsymbol{f}$ or an estimate obtained from the simulated data using the OS-EM algorithm was used in the EMG/EXP-MLE and PS methods.
The initial estimate for $\hat{\bm{\lambda}}$ was a uniform image of 0.5 ns$^{-1}$.
Given an image $\bm{x}$,
the standardized absolute log ratio (SALR) is defined as \cite{chen12enhanced}:
\begin{equation}
\text{SALR}(\bm{x})_p = \frac{ \left| \log ( {\bm x}_{p}/{\bm x}_b) \right|}{%
\text{SD}({\bm x}_b) / \overline{{\bm x}}_b},
\label{eqn:SALR}
\end{equation}
where $\overline{\bm x}$ and \(\text{SD}(\bm x)\)
are the average and standard deviation over pixels in $\bm x$,
${\bm x}_p$ contains pixels of $\bm x$ in a region of interest (ROI) $p$,
and ${\bm x}_b$ contains pixels of $\bm x$
in the surrounding background of the ROI.
As detailed in \cite{chen12enhanced}, higher SALR values
are desired because it measures the contrast of an ROI against
the background variability due to noise.
Therefore, unless mentioned otherwise,
for EMG-MLE, Exp-MLE and PS we calculate an overall SALR
by averaging \(\text{SALR}(\hat{\bm\lambda})_p\),
where $p$ indexes the disc in the phantom,
and select the iteration number that yields the largest overall SALR.
All algorithms were run with a sufficiently large number of iterations so that the iteration yielding the max SALR could be identified. 

\subsection{Evaluation}

For quantitative evaluation,
we considered the normalized mean square error (NMSE), defined as:
\begin{equation}
\label{NMSE}
\text{NMSE}=\frac{\|\hat{\bm\lambda}-{\bm\lambda}\|^2 }{ \|{\bm\lambda}\|^2},
\end{equation}
where $||\cdot||$ denotes the Euclidean norm. 
We computed pixel-wise means and standard deviations of the NMSE of 
the images reconstructed from ten independent simulations, 
using EMG-MLE, Exp-MLE, and PS with $\beta=0, 5, 10$.
Then we obtained the average mean and standard deviation inside various regions, including the four discs and the background circle excluding the discs.

We also conduct cross-correlation analysis to quantify potential cross-talk between the activity map used and 
the resulting $\hat{\bm\lambda}$.
The cross-correlation is defined as:
\begin{equation}
C:=\frac{(\hat{\bm \lambda} - \bm \lambda)^T\cdot \bm f}{\|\bm \lambda\|\|\bm f\|}.
\label{cross-correlation}
\end{equation}

\subsection{Initial examination of multi-population PLI reconstruction}
\label{subsect:two-population}
So far, we have considered the presence of only o-Ps. 
As already mentioned, substantially faster DA and p-P populations co-exist,
and proportions of the DA, p-P and o-P populations,
approximately equal to 0.6, 0.1, and 0.3, are not sensitive to the environment
\cite{moskal2021positronium}.
We extended our simulations to include two rate-constant images,
${\bm\lambda}_1$ for o-P and ${\bm\lambda}_2$ for DA,
neglecting the p-P population for now due to its relatively small population weight.
As described in Appendix, in this case $\mbox{EMG}(\tau_k;\lambda_j,\sigma^2)$ in
Eq.~(\ref{PET::profile_log_likelihood_PLI}) will be replaced with
\begin{equation}\small
w_1\text{EMG}(\tau_k;\lambda_{1,j},\sigma^2)+w_2\text{EMG}(\tau_k;\lambda_{2,j},\sigma^2),
\end{equation}
where $\lambda_{1,j}$ and $\lambda_{2,j}$ are the values of ${\bm\lambda}_1$
and ${\bm\lambda}_2$ at pixel $j$, respectively, and
$w_1$ and $w_2$ with $w_1+w_2=1$ are the population weights for o-P and DA events, respectively.
The isotope image $\bm f$ and the o-P rate-constant image ${\bm\lambda}_1$ remained
those of Phantom 2. On the other hand, the DA rate-constant image ${\bm\lambda}_2$
was the background ellipse of the phantom taking the value of 2.5\,$\mbox{ns}^{-1}$.
Also, $w_1=0.3$ was used.
The number of PLI events produced in a simulation run was again one million.
In reconstruction, the exactly known $w_1$, $\bm{\lambda}_2$ and ${\bm{f}}$
were used for obtaining the MLE of $\bm{\lambda}_1$.
More details about the two-population model can be found in \cite{chen12enhanced}.
\section{Results}

\begin{figure}[t]
\centering
\includegraphics[scale=0.64]{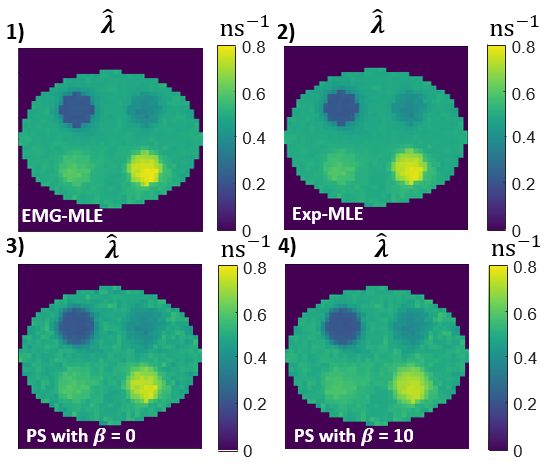}
\caption{Results obtained for Phantom 1 by using 
(1) the EMG-MLE method, (2) the Exp-MLE method,
(3) the PS method \cite{qi2022positronium} with $\beta$ = 0,
and (4) the PS method with $\beta$ = 10.
The true $\bm{f}$ is used in these reconstructions.}
\label{phantom1_MLE_surrogate}
\end{figure}

\subsection{Rate-constant image reconstruction}

We considered a CRT of 400\,ps and, as described above, 200\,ps-width TOF bins. These settings resulted in 2.52 million TOF-PET channels. 
Fig.~\ref{phantom1_MLE_surrogate} illustrates the outcomes of various reconstruction methods applied to Phantom 1. 
Fig.~\ref{phantom1_MLE_surrogate}(1) \& \ref{phantom1_MLE_surrogate}(2) show the rate-constant images, $\hat{\bm\lambda}_{\text{EMG-MLE}}$ and $\hat{\bm\lambda}_{\text{Exp-MLE}}$, 
obtained by EMG-MLE and Exp-MLE, respectively.
Fig.~\ref{phantom1_MLE_surrogate}(3) \& \ref{phantom1_MLE_surrogate}(4) show the results of the PS method by using $\beta = 0$ and $10$, respectively.
These results were obtained by using the the true activity image $\bm{f}$ and setting pixel values outside the phantom to zero.
Subjectively, the PS image using $\beta = 0$ exhibits slightly larger background variability, but overall the differences between these images are not readily apparent.

\begin{figure}[t]
\centering
\includegraphics[scale=0.65]{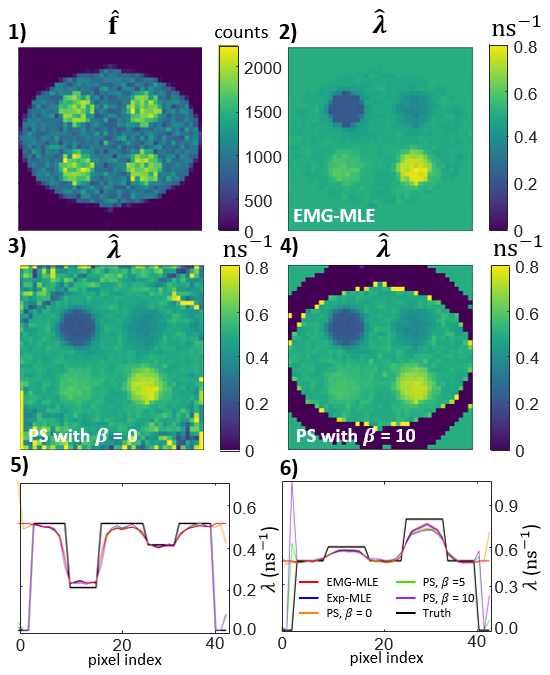}
\caption{Results obtained for Phantom 1.
$\hat{\bm f}$ in panel (1) shows the activity image
reconstructed from simulated data using OS-EM.
$\hat{\bm\lambda}$ in panels 2), 3), and 4) are rate-constant images obtained by, respectively, EMG-MLE, PS with $\beta=0$, and PS with $\beta=10$ when using $\hat{\bm f}$.
Plots in panels 5) \& 6) show horizontal profiles
across the center of, respectively, the upper and low discs of these rate-constant images.
For comparison, the true profiles and the profiles of rate-constant images obtained
by Exp-MLE and PS with $\beta=5$, using $\hat{\bm f}$, are also shown.
}
\label{phantom1_MLE}
\end{figure}

Fig.~\ref{phantom1_MLE} compares the rate-constant images obtained when using $\hat{\bm f}$ produced from the simulated data
using the OS-EM algorithm.
With $\hat{\bm f}$ and without setting the background outside the oval region as zero,
although the activity is low in the area external to the phantom,
it does not vanish.
The EMG-MLE image shown in Fig.~\ref{phantom1_MLE}(2) did not apply constraints to updating pixels outside the phantom. This indicates that the observed stable pixel values are not due to any special regularization in these external regions, where the estimated activity image value remains low.
As a result, unlike those shown in Fig.~\ref{phantom1_MLE_surrogate}, the values of the pixels external to the phantom remained close to
their initial values,
which is equal to the background circle (0.5 $\textrm{ns}^{-1}$). 
In PS, significant streaks occur when using $\beta=0$ (no regularization)
and they can be removed when using $\beta=10$.
Based on the profiles in Fig.~\ref{phantom1_MLE}(5) \& \ref{phantom1_MLE}(6), the upper discs of the EMG-MLE are comparable with, while the lower discs are closer to the true values than, those in the PS images.

\begin{figure}
\centering
\includegraphics[scale=.63]{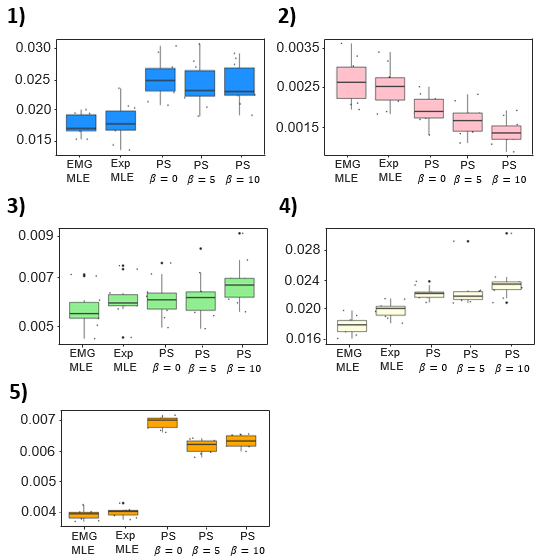}
\caption{Boxplots of NMSE of the EMG-MLE, Exp-MLE, and PS using $\beta=0, 5, 10$ reconstructions of Phantom\,1 in (1) upper-left disc, (2) upper-right disc, (3) lower-left disc, (4) lower-right disc, and (5) background disc, respectively. Simulated data contained one million events. Estimated $\hat{\bm f}$ was used in reconstruction.
}
\label{boxplot_comparisons}
\end{figure}

Fig.~\ref{boxplot_comparisons} show the NMSE results obtained for various reconstruction methods when using estimated \(\hat{\boldsymbol{f}}\).
They demonstrate that EMG-MLE can achieve lower NMSE means, or at least comparable to those of PS, particularly for the upper-left disc, the lower-right disc, and the background disc.
Most standard deviations given by EMG-MLE are lower than those by PS. 
Table~S1 in the Supplementary data additionally compares the NMSE results for various reconstruction methods when using the true $\bm{f}$ and the estimated $\hat{\bm f}$.
Notably, when using the true $\bm{f}$, EMG-MLE achieves only slightly lower 
average NMSE means and standard deviations than when using the estimated $\hat{\bm{f}}$. This suggests stability of EMG-MLE with respect to noise even though no explicitly noise regularization scheme is incorporated.

Table~\ref{crosscorr} reports the cross-correlation values calculated by using Eq.~\eqref{cross-correlation}. The low values indicate negligible interactions from $\bm{f}$ into $\hat{\bm\lambda}$ for all methods examined.
 
\begin{table*}
\begin{center}

\begin{tabular}{ |c|c|c|c|c|c|c| }  \hline	
      \multicolumn{6}{| c|}{\textbf{Cross-correlation} }\\
      \hline	 
      \multirow{2}{3em} {\textbf{Regions}} & \textbf{EMG-MLE}& \textbf{Exp-MLE}& \textbf{PS}& \textbf{PS}& \textbf{PS}\\
        & \textbf{}& \textbf{}& \textbf{$\beta$ = 0} & \textbf{$\beta$ = 5} & \textbf{$\beta$ = 10} \\

\hline
\multirow{1}{5em} {Upper\hspace{0.1cm}left} & $1.20 \times 10^{-1}$& $1.23 \times 10^{-1}$& $1.43 \times 10^{-1}$& $1.41 \times 10^{-1}$& $1.41 \times 10^{-1}$\\ 

 & $(5.91 \times 10^{-3})$& $(8.39 \times 10^{-3})$& $(8.23 \times 10^{-3})$& $(8.57 \times 10^{-3})$& $(9.02 \times 10^{-3})$\\

\hline
\multirow{2}{5em} {Upper\hspace{0.1cm}right} & $-6.14 \times 10^{-4}$& $2.64 \times 10^{-3}$& $1.11 \times 10^{-2}$& $9.45 \times 10^{-3}$& $8.13 \times 10^{-3}$\\ 

 & $(5.39 \times 10^{-3})$& $(7.58 \times 10^{-3})$& $(3.40 \times 10^{-3})$& $(4.35 \times 10^{-3})$& $(4.47 \times 10^{-3})$\\ 
 
\hline
\multirow{2}{5em} {Lower\hspace{0.1cm}left} & $-6.72 \times 10^{-2}$& $-7.08 \times 10^{-2}$& $-7.09 \times 10^{-2}$& $-7.38 \times 10^{-2}$& $-7.78 \times 10^{-2}$\\ 

 & $(5.90 \times 10^{-3})$& $(5.94 \times 10^{-3})$& $(5.41 \times 10^{-3})$& $(6.36 \times 10^{-3})$& $(5.98 \times 10^{-3})$\\ 

\hline
\multirow{2}{5em} {Lower\hspace{0.1cm}right} & $-1.20 \times 10^{-1}$& $-1.29 \times 10^{-1}$& $-1.41 \times 10^{-1}$& $-1.44 \times 10^{-1}$& $-1.49 \times 10^{-1}$\\ 

 & $(5.17 \times 10^{-3})$& $(4.07 \times 10^{-3})$& $(3.37 \times 10^{-3})$& $(8.49 \times 10^{-3})$& $(9.31 \times 10^{-3})$\\ 

\hline
\multirow{2}{5em} {Background} & $-2.73 \times 10^{-2}$& $-2.94 \times 10^{-2}$& $-3.21 \times 10^{-2}$& $-2.47 \times 10^{-2}$& $-1.53 \times 10^{-2}$\\ 

 & $(1.85 \times 10^{-3})$& $(1.62 \times 10^{-3})$& $(1.99 \times 10^{-3})$& $(1.90 \times 10^{-3})$& $(1.93 \times 10^{-3})$\\ 
\hline

\end{tabular}

\end{center}
\caption{Cross-correlations obtained for Phantom 1 using $\hat{\bm f}$. The SD is shown in the parenthesis.}
\label{crosscorr}
\end{table*}

These results underscore the efficacy of EMG-MLE in accurately reconstructing rate-constant images from TOF-PET data with a CRT of 400\,ps.

\subsection{The presence of an additional constant decay component}
\label{subsect:two-population2}

\begin{figure}
    \centering
    \includegraphics[width=0.45\textwidth]{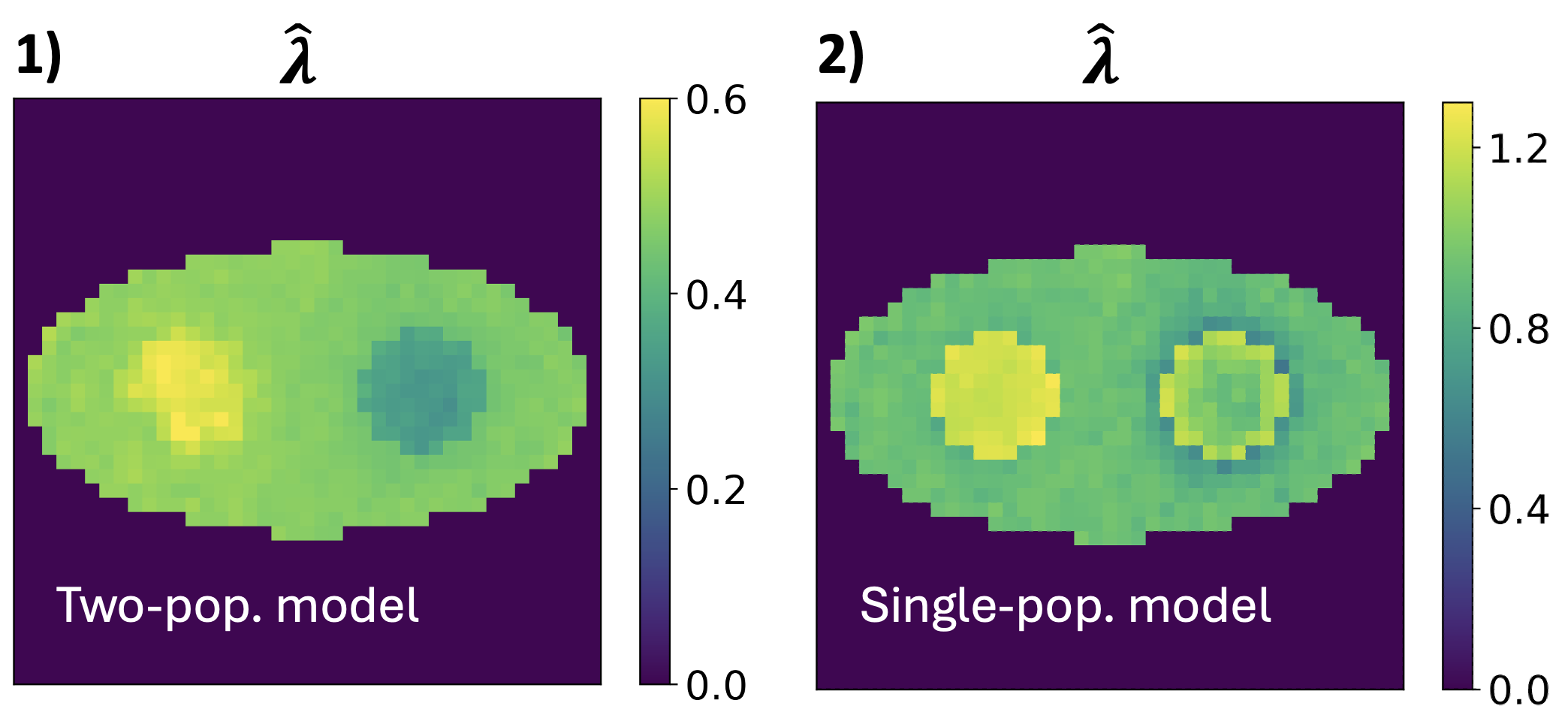}
    \caption{o-P rate-constant image obtained for Phantom 2 
    from a single simulation data consisting of two event populations
    by using (1) the single-population method and (2) two-population model.}
    \label{fig:two-pop-image}
\end{figure}

Fig.~\ref{fig:two-pop-image} shows the reconstructed o-P rate-constant image of the single-population and two-population model for Phantom 2, obtained from a single simulation, using one million events and 400\,ps CRT. The reconstructed image of the two-population model yields a higher contrast of two circular regions relative to the background compared to the single-population model. In addition, the single-population model fails to reconstruct the difference in the rate-constant value between the two circular regions. This preliminary experiment shows that the proposed method can be extended to handle multi-population data.

Table~\ref{crosscorr_two_pop} presents the cross-correlation between the activity map and the reconstructed o-Ps rate-constant images. This comparison includes results from two scenarios: one using a two-population simulation with a two-component EMG likelihood estimate, and the other using two-population simulation but with a single-component EMG likelihood estimate. Both the two-population and single-population models show negligible interaction between $\bm{f}$ and $\hat{\bm{\lambda}}$. Furthermore, the NMSE values of the reconstructed o-P rate-constant images for the three regions (left disc, right disc, and background) are summarized in Table S2 in the Supplementary data. The two-population model demonstrates significantly lower NMSE for all three regions compared to the single-population model.

\section{Summary and Discussions}
In this study, we presented a probabilistic model for finding the ML solution of the positronium lifetime image from list-mode data acquired by a TOF-PET system that is extended for detecting triple coincidences when a $\beta^{+}\gamma$ isotope such as Sc-44 is used. We conducted computer simulation studies for a 2-D TOF-PET system that emulates the configurations of human TOF-PET systems with 288 detectors on a 57\,cm diameter ring. 
CRTs of 200\,ps, 400\,ps, 600\,ps, and 800\,ps are considered for examining the effecst of CRT on the resulting lifetime images.

Our results demonstrate that the estimated rate-constant images $\hat{\bm \lambda}_\text{EMG-MLE}$ or $\hat{\bm \lambda}_\text{Exp-MLE}$ agree well with the ground truth. The differences between Exp-MLE and PS with $\beta=0$ shown in Fig.~\ref{boxplot_comparisons} may stem from the standardized absolute log ratio (SALR) stopping criterion, leading to relatively short iterations. As CRT increases, the accuracy of both methods generally decreases, but EMG-MLE consistently outperforms Exp-MLE. 
Our method incorporates Gaussian blur into the likelihood function and can be extended to a two-population model (o-P and p-P), which distinguishes it from existing approaches.

Our current studies have not accounted for attenuation, scatter, and random events.
In addition, multiple lifetime populations are present and we have only preliminary extended the proposed method to include two populations (direct annihilation of positron with
electron also is characterized by its own decay rate constant).
Moving forward, we plan to extend out model to include these factors to fit real-wold
PLI data more accurately.
Triple coincidences have a much lower probability of occurrence than the conventional annihilation coincidences. 
Although emerging systems such as the Total-Body systems can offer
an adequate sensitivity \cite{parodi2023experience}, 
we expect the PLI data to be count limited in general.
Therefore, resolution-preserving noise regularization can be an important 
consideration in PLI reconstruction.
Given the complexity, discrepancies between the model assumed for reconstruction 
and the data are very likely to exist.
Another issue of practical significance is the computation complexity
when extending the method to 3-D TOF-PET systems.
Therefore, we also plan to delve into Bayesian modeling and regularized optimization, 
coupled with parallel computation strategies, to develop accurate, robust, 
and compute-efficient reconstruction algorithms for PLI.

\begin{table*}[t]
\begin{center}

\begin{tabular}{ |c|c|c| }  \hline	
      \multicolumn{3}{| c|}{\textbf{Cross-correlation} }\\
      \hline	 
      \multirow{1}{3em} {\textbf{Regions}} & \textbf{Two-population}& \textbf{Single-population} \\
\hline
\multirow{1}{5em} {Left Disc} & $-1.15 \times 10^{-3}$& $1.67 \times 10^{-2}$ \\

\hline
\multirow{1}{5em} {Right Disc} & $-2.24 \times 10^{-3}$ & $-2.67 \times 10^{-2}$ \\
 
\hline
\multirow{1}{5em} {Background} & $-1.06 \times 10^{-4}$& $1.59 \times 10^{-3}$ \\

\hline

\end{tabular}

\end{center}
\caption{Cross-correlations obtained for the single- and two-population method for Phantom 2 from a single simulation. }
\label{crosscorr_two_pop}
\end{table*}

\section*{Appendix}

In this appendix, we elaborate on the derivation of the joint likelihood 
for the PLI data acquired by using a TOF-PET system. 
Let $f_j\geq 0$ be the PET activity in pixel $j$ so that
the number of positrons released
in pixel $j$ during an interval $T$ 
is a Poisson random number with 
mean $f_j T$.
A released positron in pixel $j$ can exist for a duration whose distribution
following an exponential distribution $\mbox{Exp}(\tau;\lambda_j)$ with some $\lambda_j\geq 0$ before annihilation.
Assume that the uncertainty in the measurement of $\tau$ is described by
$g(\tau;0,\sigma^2)$ -- a zero-mean Gaussian with standard deviation $\sigma$,
the distribution of the observed $\tau$ will be the convolution of 
$\mbox{Exp}(\tau;\lambda)$ with $g(\tau;0,\sigma^2)$,
which can be evaluated to yield the Exponentially Modified Gaussian 
$\mbox{EMG}(\tau;\lambda_j,\sigma^2)$ given in Eq.~(\ref{eq:EMG}).
Subsequently, annihilation of a positron produces two opposite 511\,keV photons
that may be detected by the TOF-PET system
at an LOS, say $c=(i_{511keV,1},i_{511keV,2},m)$
where $i_{511keV,1}$ and $i_{511keV,2}$ are the indices of the detectors making
the detection and $m$ is the TOF index (the TOF value is typically binned).
In this treatment, a PLI detection is given by $(c,\tau)$.

Let $\bm{f}=[f_j]^t$ and $\bm\lambda=[\lambda_j]^t$
be vectors representing the activity and rate-constant images.
Given a PLI detection, we denote by $p(c, \tau| \bm{\lambda}, \bm{f})d\tau$
the probability for it to be in the set 
$S_{d\tau}(c,\tau)=\{(c,\tau'):\tau\leq\tau'<\tau+d\tau\}$
when the activity and rate-constant image
are $\bm{f}$ and $\bm\lambda$, respectively.
Evidently, $p(c, \tau| \bm{\lambda}, \bm{f})d\tau$
equals to the ratio of $n_1(S_{d\tau}(c,\tau))$, 
the expected number of detections in $S_{d\tau}(c,\tau)$,
to $n_2$, the expected total number of all detections.
Evidently, 
\begin{align}
n_1(S_{d\tau}(c,\tau)) = \sum_{j=1}^{N_j} f_j T \times \mbox{EMG}(\tau;\lambda_j,\sigma^2)d\tau \times H_{c,j}, 
\label{mean_given_jk}
\end{align}
where $f_j T$ gives the expected number of positrons in pixel $j$,
$\mbox{EMG}(\tau;\lambda_j,\sigma^2)d\tau$ gives the probability for one of these positrons 
to exist for a duration $\tau'\in[\tau,\tau+d\tau)$,
and $H_{c,j}$ gives the probability for an annihilation in pixel $j$
to yield a detection at LOS $c$.
Here we have made two assumptions: 
(i) the isotope decay, positron annihilation, and coincidence detection in a chain
leading to a PLI detection are independent processes, and
(ii) the processes involved in different PLI detections are independent.
$n_2$ is the totality of $n_1(S_{d\tau}(c,\tau))$ over all possible $c$ and $\tau$:
\begin{equation}
n_2 = \sum_c \int_{\tau}S_{d\tau}(c,\tau)d\tau
=T s_{\bm f},
\end{equation}
where $s_{\bm f} = \sum_{c,j} H_{c,j} f_j$.
Therefore,
\begin{equation}
p(c, \tau | \bm{\lambda}, \bm{f}) = \frac{1}{d\tau}\frac{n_1}{n_2} 
= \frac{\sum_{j=1}^{N_j}H_{c,j} f_j\ \mbox{EMG}(\tau;\lambda_j,\sigma^2)}{s_{\bm f}}.
\end{equation}
Note that $p(c, \tau| \bm{\lambda}, \bm{f})$ 
is a probability in $c$ but a density in $\tau$.

Under independence assumptions,
given $N_k$ detections the probability for them to be $(c_1,\tau_1),\cdots,(c_{N_k},\tau_{N_k})$
is therefore $p(\mathcal{W}_{N_k}^{0};\bm{\lambda},\bm{f})=\prod_{k=1}^{N_k} p(c_k, \tau_k | \bm{\lambda}, \bm{f})$, where $\mathcal{W}_{N_k}^{0}=\{(c_k,\tau_k):1\leq k \leq N_k\}$.
Consequently, we have the log-likelihood
\begin{equation}\footnotesize
\ell(\bm{\lambda},\bm{f};\mathcal{W}_{N_k}^{0})= 
\sum_{k=1}^{N_k} \log \left( \sum_{j=1}^{N_p} H_{c_k,j}f_j\,\mbox{EMG}(\tau_k;\lambda_j, \sigma^2) \right)
- N_k \log(s_{\bm f}).
\label{eq:joint-loglike-PC}
\end{equation}

Above we consider preset-count (PC) acquisition in which a scan is stopped
when reaching a prescribed number $N_k$ of detection.
Alternatively, in  preset-time (PT) acquisition a scan is conducted for 
a prescribed duration $T$ and hence $N_k$ becomes a Poisson random number
with mean $n_2=T s_{\bm f}$.
With PT acquisition, the probability for having $N_k$ detections and
they are given by $\mathcal{W}_{N_k}^{0}$ is then
$p'(\mathcal{W}^0_{N_k};\bm{f},\bm{\lambda},T)
= \text{Poisson}(N_k;T s_{\bm f}) \times p(\mathcal{W}_{N_k}^{0};\bm{\lambda},\bm{f}),$
yielding the log-likelihood (omitting constants):
\begin{equation}\footnotesize
\ell'(\bm{\lambda},\bm{f};\mathcal{W}^0_{N_k},T)
= \sum_{k=1}^{N_k} \log \left( \sum_{j=1}^{N_j} H_{c_k,j}f_j\,\mbox{EMG}(\tau_k;\lambda_j,\sigma^2) \right) - T s_{\bm f}.
\label{eq:joint-loglik-PT}
\end{equation}
If $\bm{f}$ is known \textit{a priori}, the 2nd terms of these log-likelihoods
are constants. Therefore, omitting constants both acquisition modes
have the same profile log-likelihood given $\bm{f}$:
\begin{equation}
\ell(\bm{\lambda};\mathcal{W}^0_{N_k},\bm{f})
= \sum_{k=1}^{N_k} \log \left( \sum_{j=1}^{N_j} H_{c_k,j}f_j\,\mbox{EMG}(\tau_k;\lambda_j,\sigma^2) \right).
\label{eq:profile_loglik_PT}
\end{equation}
This profile likelihood function extends the model in \cite{qi2022positronium} by replacing the exponential distribution with the exponentially modified Gaussian (EMG) distribution.

The gradient of $\ell(\bm{\lambda};\mathcal{W}^0_{N_k},\bm{f})$
is provided to the L-BFGS-B algorithm to speed up computation.
It is an exercise to show
\begin{equation}
\frac{\partial \ell(\bm{\lambda};\bm{f};\mathcal{W}^0_{N_k})}{\partial \lambda_j}
= \sum_{k=1}^{N_k} \frac{H_{c_k,j}f_j\,\mbox{dEMG}(\tau_k;\lambda_j,\sigma^2)}
{\sum_{j'=1}^{N_j} H_{c_k,j'} f_{j'}\,\mbox{EMG}(\tau_k;\lambda_{j'},\sigma^2)},
\end{equation}
where
\begin{align}
\text{dEMG}&(\tau;\lambda,\sigma^2) =
\partial\text{EMG}(\tau;\lambda,\sigma^2)/\partial\lambda \nonumber\\
=& \frac12 e^{-\lambda (\tau-\frac12\sigma^2\lambda)}
    \left\{ -\frac{\sigma\lambda}{\sqrt{2\pi}}
        \exp^{-(\tau-\sigma^2\lambda)^2/2\sigma^2}
    \right. \nonumber\\
&+ \left. \left(1-\lambda\tau+\sigma^2\lambda^2\right)
        \left( 1 + \mbox{erf}\left( 
                        \frac{\tau-\lambda\sigma^2}{\sqrt{2}\sigma}\right)
        \right) 
    \right\}.
\end{align}

As described in \cite{moskal2019positronium}, positrons in fact can exist
in multiple different states before annihilation, each of which is
characterized by a distinct rate constant.
In general, multiple rate-constant images, say
${\bm\lambda}_p,\ p=1,\ldots,N_p$, are therefore needed,
and the probability distribution for the observed $\tau$ becomes
$\sum_{p} w_p \mbox{EMG}(\tau;\lambda_{p,j},\sigma^2)$ where
$w_p$'s with $\sum_p w_p=1,\,w_p\geq0$ are the proportions of
positrons existing in respective "populations" (we shall assume that
the population weights do not vary over pixels) and
$\lambda_{p,j}$ is the value of $\bm{\lambda}_p$ at pixel $j$.
Following the above derivation, the profile log-likelihood 
given $\bm{f}$ is
\begin{align}\small
\ell(&w_1, \bm{\lambda}_1, \ldots, w_{N_p}, \bm{\lambda}_{N_p};
\mathcal{W}^0_{N_k},\bm{f}) \nonumber \\
&= \sum_{k=1}^{N_k} \log \left( \sum_{j=1}^{N_j} H_{c_k,j}f_j\,
\sum_{p=1}^{N_p} w_p\, \mbox{EMG}(\tau_k;\lambda_{p,j}, \sigma^2) \right).
\end{align}
In Section~\ref{subsect:two-population}, we consider the case with $N_p=2$
and treat $w_1$, $w_2$, and ${\bm\lambda}_{2}$ as constants,
obtaining $\ell(\bm{\lambda}_1;w_1,w_2,\bm{\lambda}_2,\mathcal{W}^0_{N_k},\bm{f})$.
In this case, we need the gradient
\begin{align}
\partial\ell(\bm{\lambda}_1; &w_1,w_2,\bm{\lambda}_2,\mathcal{W}^0_{N_k},\bm{f})/
\partial \lambda_{1,j} \nonumber\\
&= \sum_k \frac{H_{c_k,j}f_j w_1\,\mbox{dEMG}(\tau_k;\lambda_{1,j},\sigma^2)}%
{\sum_{j'} H_{c_k,j'} f_{j'}\,\sum_{p=1}^2 w_p\,\mbox{EMG}(\tau_k;\lambda_{p,j'},\sigma^2)}.
\end{align}

Although this paper does not consider the simultaneous MLE of
$\bm{f}$ and $\bm{\lambda}$ using the joint log-likelihoods,
several observations are worth noting.
First, from Eq.~(\ref{eq:joint-loglike-PC}) it can be readily checked that
$\ell(\bm\lambda,\xi\bm{f};\mathcal{W}_{N_k}^{0})
=\ell(\bm\lambda,\bm{f};\mathcal{W}_{N_k}^{0})$ for any $\xi>0$.
This reflects that, since PC acquisition does not care about scan time, $\bm{f}$ scanned for an interval $\xi T$ and
$\xi\bm{f}$ scanned for an interval $T$ are equally likely to
explain a given dataset.
This observation leads us to stipulate that the log-likelihood
maximizing $\bm{f}$ for PC- and PT-acquisitions differ only in scale.
To verify this, using Eq.~(\ref{eq:joint-loglike-PC}) \& (\ref{eq:joint-loglik-PT})
we have
\begin{equation}
\ell'(\bm{\lambda},\xi\bm{f};\mathcal{W}_{N_k}^{0})
= \ell(\bm{\lambda},\bm{f};\mathcal{W}_{N_k}^{0}) + \alpha(\bm{f},\xi),
\label{eq:PT-PC}
\end{equation}
where $\alpha(\bm{f},\xi) = N_k\log(s_{\bm{f}})+N_k\log(\xi)-\xi T\,s_{\bm{f}}$.
One can easily check that, for any $\bm{f}$,
$\alpha(\bm{f},\xi)$ can reach its  maximum value of $N_k\log(N_k/eT)$
by setting $\xi=N_k/Ts_{\bm{f}}$.
Therefore, if $(\bm{f}^*,\bm{\lambda}^*)$ maximizes
$\ell(\bm{\lambda},\bm{f};\mathcal{W}_{N_k}^{0})$, then
$(\xi^*\bm{f}^*,\bm{\lambda}^*)$, $\xi^*=N_k/Ts_{\bm{f}^*}$
maximizes $\ell'(\bm{\lambda},\bm{f};\mathcal{W}_{N_k}^{0})$
because both terms on the right-hand side of Eq.~(\ref{eq:PT-PC}) are maximized.
Consequently, it is sufficient to consider only the PC joint log-likelihood $\ell(\bm{\lambda},\bm{f};\mathcal{W}_{N_k}^{0})$.
It is not difficult to see that the role of scaling by $\xi^*$
is to fix the activity of a PC MLE solution to produce a PT MLE solution
that best accounts for the observation of $N_k$ events during the prescribed
scan time $T$.

\bibliographystyle{IEEEtran}  
\bibliography{FINAL_VERSION}

\renewcommand{\thetable}{S1}


\begin{table*}[h]
\begin{center}

\begin{tabular}{ |c|c|c|c|c|c|c|c| }  \hline	
      \multicolumn{7}{| c|}{\textbf{Mean (Standard deviation) of NMSE} from the 10 simulation replicates }\\
      \hline	 
      \multirow{2}{3em} {\textbf{Given}} & \multirow{2}{3em} {\textbf{Regions}} & \textbf{Proposed}& \multirow{2}{5em} {\textbf{Exp-MLE}}& \textbf{PS}& \textbf{PS}& \textbf{PS}\\
        & & \textbf{method}& \textbf{}& \textbf{$\beta$ = 0} & \textbf{$\beta$ = 5} & \textbf{$\beta$ = 10} \\

\hline
\multirow{9}{1em} {$\bm{f}$} & \multirow{2}{5em} {Upper\hspace{0.1cm}left} & $1.88 \times 10^{-2}$& $1.88 \times 10^{-2}$& $2.86 \times 10^{-2}$& $2.64 \times 10^{-2}$& $2.56 \times 10^{-2}$\\ 

 & & $(1.93 \times 10^{-3})$& $(1.78 \times 10^{-3})$& $(3.35 \times 10^{-3})$& $(2.70 \times 10^{-3})$& $(2.99 \times 10^{-3})$\\

\cline{2-7}
 & \multirow{2}{5em} {Upper\hspace{0.1cm}right} & $1.93 \times 10^{-3}$& $1.79 \times 10^{-3}$& $1.93 \times 10^{-3}$& $1.64 \times 10^{-3}$& $1.30 \times 10^{-3}$\\ 

 & & $(3.70 \times 10^{-4})$& $(2.67 \times 10^{-4})$& $(3.48 \times 10^{-4})$& $(3.62 \times 10^{-4})$& $(3.23 \times 10^{-4})$\\ 
 
\cline{2-7}
& \multirow{2}{5em} {Lower\hspace{0.1cm}left} & $5.43 \times 10^{-3}$& $6.15 \times 10^{-3}$& $6.32 \times 10^{-3}$& $6.40 \times 10^{-3}$& $6.99 \times 10^{-3}$\\ 

 & & $(7.64 \times 10^{-4})$& $(8.90 \times 10^{-4})$& $(8.33 \times 10^{-4})$& $(8.83 \times 10^{-4})$& $(8.87 \times 10^{-4})$\\ 

\cline{2-7}
 & \multirow{2}{5em} {Lower\hspace{0.1cm}right} & $1.61 \times 10^{-2}$& $1.86 \times 10^{-2}$& $2.31 \times 10^{-2}$& $2.27 \times 10^{-2}$& $2.34 \times 10^{-2}$\\ 

& & $(9.54 \times 10^{-4})$& $(1.04 \times 10^{-3})$& $(8.67 \times 10^{-4})$& $(1.22 \times 10^{-3})$& $(1.33 \times 10^{-3})$\\ 

\cline{2-7}
 & \multirow{2}{5em} {Background} & $2.99 \times 10^{-3}$& $3.09 \times 10^{-3}$& $6.60 \times 10^{-3}$& $5.66 \times 10^{-3}$& $5.62 \times 10^{-3}$\\ 

 & & $(1.77 \times 10^{-4})$& $(1.82 \times 10^{-4})$& $(1.95 \times 10^{-4})$& $(2.07 \times 10^{-4})$& $(1.98 \times 10^{-4})$\\ 
\hline

\multirow{9}{1em} {$\hat{\bm{f}}$} & \multirow{2}{5em} {Upper\hspace{0.1cm}left} & $1.75 \times 10^{-2}$& $1.80 \times 10^{-2}$& $2.52 \times 10^{-2}$& $2.42 \times 10^{-2}$& $2.42 \times 10^{-2}$\\ 

 & & $(1.78 \times 10^{-3})$& $(3.03 \times 10^{-3})$& $(3.25 \times 10^{-3})$& $(3.62 \times 10^{-3})$& $(3.37 \times 10^{-3})$\\

\cline{2-7}
 & \multirow{2}{5em} {Upper\hspace{0.1cm}right} & $2.59 \times 10^{-3}$& $2.43 \times 10^{-3}$& $1.88 \times 10^{-3}$& $1.62 \times 10^{-3}$& $1.33 \times 10^{-3}$\\ 

 & & $(5.23 \times 10^{-4})$& $(4.87 \times 10^{-4})$& $(3.46 \times 10^{-4})$& $(3.66 \times 10^{-4})$& $(3.07 \times 10^{-4})$\\ 
 
\cline{2-7}
& \multirow{2}{5em} {Lower\hspace{0.1cm}left} & $5.78 \times 10^{-3}$& $6.18 \times 10^{-3}$& $6.20 \times 10^{-3}$& $6.30 \times 10^{-3}$& $6.88 \times 10^{-3}$\\ 

 & & $(8.52 \times 10^{-4})$& $(8.53 \times 10^{-4})$& $(8.02 \times 10^{-4})$& $(9.49 \times 10^{-4})$& $(9.49 \times 10^{-4})$\\ 

\cline{2-7}
 & \multirow{2}{5em} {Lower\hspace{0.1cm}right} & $1.79 \times 10^{-2}$& $2.01 \times 10^{-2}$& $2.25 \times 10^{-2}$& $2.28 \times 10^{-2}$& $2.40 \times 10^{-2}$\\ 

& & $(1.26 \times 10^{-3})$& $(1.13 \times 10^{-3})$& $(8.93 \times 10^{-4})$& $(2.58 \times 10^{-3})$& $(2.70 \times 10^{-3})$\\ 

\cline{2-7}
 & \multirow{2}{5em} {Background} & $3.55 \times 10^{-3}$& $3.63 \times 10^{-3}$& $6.70 \times 10^{-3}$& $5.89 \times 10^{-3}$& $6.05 \times 10^{-3}$\\ 

 & & $(1.68 \times 10^{-4})$& $(1.59 \times 10^{-4})$& $(2.14 \times 10^{-4})$& $(2.29 \times 10^{-4})$& $(2.06 \times 10^{-4})$\\ 
 \hline

\end{tabular}

\end{center}
\caption{Comparisons of NMSE using the five methods for Phantom 1 with $\bm f$ and $\hat{\bm f}$. NMSE measurements of reconstructed images were taken across various regions including the upper left, upper right, lower left, lower right discs, and the background area defined by an ellipse excluding the four discs.}
\label{NMSE}
\end{table*}

\renewcommand{\thetable}{S2}

\begin{table*}[t]
\begin{center}

\begin{tabular}{ |c|c|c| }  \hline	
      \multicolumn{3}{| c|}{\textbf{NMSE}  from a single simulation}\\
      \hline	 
      \multirow{1}{3em} {\textbf{Regions}} & \textbf{Two-population}& \textbf{Single-population} \\
\hline
\multirow{1}{5em} {Left Disc} & $7.13 \times 10^{-3}$& $1.38$ \\

\hline
\multirow{1}{5em} {Right Disc} & $2.07 \times 10^{-2}$ & $2.71$ \\
 
\hline
\multirow{1}{5em} {Background} & $4.06 \times 10^{-3}$& $0.66$ \\

\hline

\end{tabular}

\end{center}
\caption{Comparisons of NMSE using the single- and two-population method for Phantom 2 from a single simulation. }
\label{NMSE_two_pop}
\end{table*}


\end{document}